\documentclass[amsmath,amssymb,aps,prd,reprint,twocolumn,showpacs,nofootinbib]{revtex4-1}
\usepackage{graphicx}

\usepackage[colorlinks=true,linkcolor=blue,citecolor=blue,urlcolor=blue]{hyperref}

\newcommand{\dd}{{d}}
\newcommand{\ee}{{e}}
\newcommand{\ii}{{i}}

\newcommand{\pdx}{\partial_x}
\newcommand{\pdt}{\partial_t}

\newcommand{\dx}{\dd x}
\newcommand{\dom}{\dd\omega}

\newcommand{\php}[1]{\ee^{+\ii #1}}
\newcommand{\phm}[1]{\ee^{-\ii #1}}

\newcommand{\chor}{{c_0}}

\newcommand{\ka}{{\kappa}}
\newcommand{\kk}{{\kappa_\alpha}}

\newcommand{\Th}{{T_{\rm H}}}

\newcommand{\Teff}{T_{\rm eff}}
\newcommand{\Tstep}{T_{\rm step}}
\newcommand{\Tav}{T_{\rm av}}

\newcommand{\Tveff}{T_{\rm eff}^v}
\newcommand{\Tvstep}{T_{\rm step}^v}

\newcommand{\kone}{{k_{-\om}^{\rm in}}}
\newcommand{\ktwo}{{k_{-\om}^{\rm out}}}

\newcommand{\om}{\omega}

\newcommand{\omm}{\omega_{\rm max}}
\newcommand{\ommin}{\omega_{\rm min}}
\newcommand{\ommax}{\omm}
\newcommand{\ommok}{\omega_{\rm max}/\ka}

\newcommand{\pom}{\phi_\omega^\alpha(x)}

\newcommand{\vpom}{\varphi_\omega^\alpha(x)}
\newcommand{\vpoms}{(\varphi_\omega^{\alpha}(x))^*}

\newcommand{\aom}{\hat a_\omega^\alpha}
\newcommand{\aomd}{\hat a_\omega^{\alpha\dagger}}

\newcommand{\hP}{\hat\Psi}
\newcommand{\Pn}{\Psi_0}
\newcommand{\rn}{\rho_0}

\newcommand{\hp}{\hat\phi}

\newcommand{\hpd}{\hat\phi^\dagger}

\newcommand{\ha}{\hat a}

\newcommand{\cd}{c_d}
\newcommand{\vd}{v_d}
\newcommand{\cu}{c_u}
\newcommand{\vu}{v_u}

\newcommand{\cn}{c_0}

\newcommand{\thp}{\theta(x)}
\newcommand{\thm}{\theta(-x)}

\newcommand{\nstep}{n_\om^{\rm step}}
\newcommand{\nvstep}{n_\om^{v, \, \rm step}}

\begin{document}

\title{Hawking radiation in dispersive theories, the two regimes}

\author{Stefano Finazzi}
\email{finazzi@science.unitn.it}
\affiliation{INO-CNR BEC Center and Dipartimento di Fisica, Universit\`a di Trento, via Sommarive 14, 38123 Povo---Trento, Italy}
\author{Renaud Parentani}
\email{renaud.parentani@th.u-psud.fr}
\affiliation{Laboratoire de Physique Th\'eorique, CNRS UMR 8627, B{\^{a}}timent 210, 
\\ Universit\'e Paris-Sud 11, 91405 Orsay Cedex, France}

\begin{abstract}
We compute the black hole radiation spectrum in the presence of high-frequency dispersion in a large set of situations. 
In all cases, the spectrum diverges like the inverse of the Killing frequency.
When studying the low-frequency spectrum, we find only two regimes: an adiabatic one where the corrections with respect to the standard temperature are small, and an abrupt one 
regulated by dispersion, in which the near-horizon metric can be replaced by step functions.
The transition from one regime to the other is governed by a single parameter which also governs the net redshift undergone by dispersive modes.
These results can be used to characterize the quasiparticles spectrum of recent and future experiments aiming to detect 
the analogue Hawking radiation.
They also apply to theories of quantum gravity which violate Lorentz invariance.
\end{abstract}

\date{\today}
\maketitle

\section{Introduction}
\label{sec:intro}

For long wavelengths, the propagation of sound waves in a moving fluid is analogous to that of light in a curved metric~\cite{Unruh81}. 
However, Hawking radiation issues from very short wavelength modes~\cite{Unruh7X?,TJ91,93,Primer}. 
Therefore, dispersion effects must be taken into account when computing the spectrum emitted by an acoustic black hole. 
When the dispersive scale $\xi$ (the healing length in a Bose condensate) is much smaller than the surface gravity scale $1/\ka$, which fixes the Hawking temperature $\Th = \ka/2\pi$ (in units where $\hbar = k_{\rm B} = c = 1$), it is now well-understood that the spectrum is robust~\cite{Unruh95,BMPS95,CJ96,Corley97,Tanaka99,UnruhSchu05,Rivista,MacherRP1,MacherBEC,broad_hor,granada_proc,cpf}.
However, the precise extension of this regime, 
and the spectral properties outside it, are still poorly understood.

The aim of this paper is to remedy this situation.
Doing so, we shall complete Ref.~\cite{broad_hor} where we studied flow profiles with a local perturbation which strongly varies close to the horizon.
Up to a certain high frequency introduced by dispersion, it was observed that the spectrum remains accurately Planckian even when the temperature significantly differs from the standard one. 
Hence, below that frequency, the whole spectrum is accurately characterized by the low-frequency effective temperature. 
Analyzing this temperature, two regimes were found: as long as the perturbation scale is larger than a certain critical length, the temperature remains very close to the standard one fixed by $\kappa$. 
For variations on shorter distances instead, the temperature is given by a spatial average over that critical length.
Interestingly, the latter is not simply given by the dispersive length $\xi$. 
Rather, it scales as $\xi^{2/3}\, \ka^{-1/3}$ for quartic dispersion.

To complete that analysis, we determine the low-frequency temperature in simpler flows characterized by fewer parameters.
We observe two well-defined regimes.
First, when the flow is smooth and dispersion negligible, the spectrum slightly differs from the relativistic one. 
Second, there is an opposite regime which is well-described by using steplike profiles to characterize the near-horizon metric~\cite{Recati2009,carlos}.
With numerical techniques~\cite{MacherBEC,broad_hor,granada_proc}, we investigate the intermediate situations and show that there are no other regimes than these two, in agreement with Ref.~\cite{Scott_thesis}.
We also show that the transition from one regime to the other is governed by a single parameter which scales in the same way as the above-mentioned critical distance.
Interestingly, this parameter also fixes the (finite) redshift undergone by dispersive modes when scattered on a sonic horizon. 
For definiteness, we consider the phonon spectra in the context of atomic Bose condensates. Yet, our analysis applies to other superluminal dispersion relations, as well as subluminal ones, in virtue of the correspondence~\cite{cpf} between the various types of dispersion relations.
It can therefore be used to explore the consequences on black hole thermodynamics~\cite{Ted_wall_review} which arise from violations of Lorentz invariance found in certain theories of quantum gravity~\cite{Ted_EinstAether,Parentani2001,HoravaPRD79,Blas_Sibiryakov2011}.

The paper is organized as follows. We first present the relevant parameters which govern the mode scattering on a sonic horizon.
We then compute the low-frequency limit of the emitted fluxes in steplike flow profiles.
Finally, by numerically scanning a wide parameter space, we characterize the validity domains of both the Hawking and the steplike regimes. 
In the appendices, we present the basic concepts governing phonon fluxes, as well as the tools needed to compute these fluxes in steplike flows.
We also show that our main results remain unchanged when considering a larger class of flows.

\section{Relevant concepts}
\label{sec:param}

\subsection{Number of $e$-folds and critical frequency $\ommax$}
\label{subsec:ommax}

We consider elongated condensates stationary flowing along the longitudinal direction $x$, and we assume that the transverse dimensions are small enough that phonon excitations are longitudinal. 
In the hydrodynamic regime, phonon propagation is governed by the d'Alembert equation in the metric~\cite{Unruh81,lr} 
\begin{equation}\label{eq:metric}
 \dd s^2=-c^2 \dd t^2+(\dd x-v \, \dd t)^2~,
\end{equation}
where the flow velocity $v$ and the speed of sound  $c$ only depend on $x$.
We assume that the condensate flows from right to left ($v<0$) and that the sonic horizon (where $w = c+v$ crosses $0$) is located at $x=0$. When the gradient
\begin{equation}\label{eq:temprelnew}
\ka \equiv \partial_x w \vert_{x = 0}
\end{equation}
is positive, one obtains an analogue black hole horizon. Indeed, in the near-horizon region (NHR) where $w \sim \ka x $ is a good approximation, the null outgoing (upstream) geodesics are $x = x_0 \, e^{ \ka (t - t_0)}$.
Correspondingly, the wave vectors obey~\cite{MisnerThorneWheeler,Primer}
\begin{equation}\label{eq:ktdep}
k(t) = k_0\,  e^{- \ka (t - t_0)}.
\end{equation}
These equations are the signatures of a (Killing) horizon. 
Notice that this exponential decay law is also found in an inflationary 
de~Sitter space where the Hubble parameter $H$ plays the role of $\ka$.
In that case, the number of $e$-folds $N= H (t - t_0) $ is used to quantify the duration of the inflationary period. The origin of this instructive correspondence is simple:
When $H = \kappa$, the de~Sitter metric in Lema\^{i}tre coordinates reads $ds^2 = -dt^2 + (dX - \kappa X dt)^2$. 
To leading order in $\kappa x \ll 1$, where $x=X + 1/\kappa$, it coincides with the metric of Eq.~(\ref{eq:metric}) evaluated in the NHR when imposing that $\chor$, the speed of sound at the horizon, is 1. For more details, see Sec.~3.3 in Ref.~\cite{Constructing}.

When ignoring high-frequency dispersion, Eq.~\eqref{eq:ktdep} implies that $n_\om$, the spectrum of upstream phonons spontaneously emitted from the horizon, is Planckian with Hawking temperature $\Th = \ka/ 2 \pi $, in units where $\hbar = k_B = 1$.
When introducing dispersion for frequencies above $\Lambda \gg \kappa$, the characteristics of the wave equation no longer follow null geodesics close to the horizon~\cite{BMPS95}.
Nevertheless, Eq.~\eqref{eq:ktdep} still applies in the NHR, and this is the root of the robustness of the spectrum~\cite{Rivista,cpf}.
To characterize the properties of $n_\om$, we define the temperature function $T_\om$ by
\begin{equation}\label{eq:Tom}
 n_\om \equiv \frac{1}{\exp(\om/ T_\om)-1}.
\end{equation}
As such, $T_\om$ is simply another way to express $n_\om$. 
It is particularly useful because it was observed~\cite{MacherBEC,broad_hor} that $n_\om$ stays accurately Planckian (see also Fig.~\ref{fig:tom} of Appendix~\ref{app:D2}) over a wide range of frequencies even when the temperature significantly differs from the relativistic one. 
In addition, in strongly dispersive regimes, i.e. for $\Lambda/\kappa < 1$, it was found that $n_\om$ still diverges like $1/\om$ for $\om \to 0$~\cite{MacherBEC,Recati2009}. 
Hence, in all cases, the low-frequency part of the spectrum is governed by the effective temperature
\begin{equation}\label{eq:defT}
 \Teff 
\equiv\lim_{\om\to0}T_\om=\lim_{\om\to0}\om\, n_\om.
\end{equation}
We define the {\it Hawking regime} by situations where $\Teff$ differs from  $\Th$ 
by less than 10\%. We aim to determine the extension of this regime when introducing short-distance dispersion.

In a stationary flowing atomic Bose condensate, the dispersion relation of phonons is given by
\begin{equation}
\label{eq:dispersion_hom}
(\om-vk)^2=\Omega^2(k)
= c^2 k^2 + \frac{\hbar^2 k^4}{4 m^2 } = c^2 k^2 +  \frac{\chor^4 k^4}{\Lambda^2},
\end{equation}
where $\om$ is the conserved frequency $i\partial_t$, $k$ the wave vector $-i\partial_x$, $m$ the atom mass, $\chor$ the speed of sound at the horizon, and $\Lambda$ the frequency associated with the healing length $\xi = \hbar/\sqrt{2}m\chor$ by $\Lambda = \sqrt{2}\chor/\xi$.
This dispersion relation directly follows from the Bogoliubov-de~Gennes equation~\cite{DalfovoRMP}, see Eq.~\eqref{eq:system} in Appendix~\ref{app:system}, where more details can be found.

Because of dispersion, as mentioned above, upstream wave packets no longer follow null geodesics. Yet, when they propagate in the NHR, the exponential decay law of Eq.~\eqref{eq:ktdep} is still satisfied when $\kappa/\Lambda \ll 1$ and $\om/\Lambda\ll 1$. 
But what is essential here is that the trajectories stay in the NHR only for a finite time, even when the above inequalities are not satisfied. As a result, unlike what is found in de~Sitter space, the accumulated redshift effect saturates at a finite value (when the asymptotic values of $c$ and $v$ are finite, which is the case we shall consider).
To quantify this saturation, we use
\begin{equation}\label{eq:rom}
e^{N_\om} = 
 r_\om \equiv \frac{\kone}{\ktwo} ,
\end{equation}
where $\kone$ and $\ktwo$ are the asymptotic values of the ingoing and outgoing wave numbers, solutions of Eq.~\eqref{eq:dispersion_hom} with $\Omega<0$, see Appendix~\ref{app:system}.
These roots describe the negative frequency partner which is trapped in the supersonic region, on the left of the horizon.
As we shall see, the number of e-folds $N_\om$, or its exponential $r_\om$, offers a simple description of the low-frequency spectrum outside the Hawking regime. It is thus important to determine how it depends on the dispersive scale $\Lambda$ and on the flow profile parameters.

To reduce the number of these parameters, we impose
\begin{equation}\label{eq:cv}
 c(x)=\chor+(1-q) w(x),\quad
 v(x)=-\chor+q \, w(x) .
\end{equation}
By introducing the parameter $q$, we specify how $w = c+v$ is shared between $c$ and $v$. 
In the forthcoming numerical simulations, we shall not use special values of $q$, such as 0 (constant flow velocity) or 1 (constant speed of sound), in order to obtain results which are generically valid.
Notice that $q=0.5$ is also particular as it minimizes the coupling between right- and left-going waves~\cite{MacherBEC}. 
As a representative value of the generic case, we choose $q=0.3$. For more details on the impact of $q$ on the spectrum, we refer to Ref.~\cite{MacherBEC}.
The flow profile is taken to be\footnote{In Appendix~\ref{app:D2}, we consider more general profiles and observe that the main results are not significantly modified.} 
\begin{equation}\label{eq:velocity}
 \frac{w(x)}{\chor} = D \tanh \left(\frac{\ka x}{\chor D}\right).
\end{equation}
The important quantity introduced here is $D$. 
It fixes the asymptotic value of $w$, but, more importantly, it also fixes the extension of the NHR, i.e., the domain where $w \sim \ka x$. 
This latter fact renders $D$ a very relevant spectral parameter in the presence of dispersion~\cite{broad_hor,cpf}.

To compute $r_\om$ of Eq.~\eqref{eq:rom}, we need to analyze the two roots of Eq.~\eqref{eq:dispersion_hom} with $\Omega < 0$. In the supersonic region, these roots merge in a point that corresponds to $x^{\rm tp}(\om)$, the turning point of the trajectory at fixed $\om$~\cite{cpf}.
Its location is fixed by solving $v_{\rm gr} = 0$, where the group velocity (in the horizon rest frame) is
\begin{equation}\label{eq:gr_vel}
v_{\rm gr} = 
\partial_k \omega =
\partial_k (\Omega + vk). 
\end{equation}
Using Eq.~\eqref{eq:dispersion_hom}, for $\om \ll \Lambda$, one finds~\cite{CJ96}
\begin{equation}
\label{eq:tp}
\frac{\ka  x^{\rm tp}(\om)}{c_0} \approx - \left(\frac{\om}{\Lambda}\right)^{2/3}.
\end{equation}
For $\om/\Lambda \to 0$, $x^{\rm tp}$ coincides with the horizon at $x=0$, as expected from the behavior of the trajectories in the absence of dispersion.
When $\om$ increases, the turning point recedes to the left. 
We define the critical frequency $\omm$ by the value of $\om$ where $x^{\rm tp}(\om)$ reaches $-\infty$.
In the velocity flows of Eq.~\eqref{eq:velocity}, one has $\omm = \Lambda \, f(D,q)$ where $f$ is a rather complicated function.
For small values of $D$, which will be a relevant regime in the sequel, one gets
\begin{equation}
\label{eq:ommax}
 \omm=\Lambda \left(\frac{2}{3}\,D\right)^{3/2}\left[1+\frac{D}{2}\left(q-\frac{5}{6}\right)+O\left(D^{2}\right)\right].
\end{equation}
Using the asymptotic values of $v$ and $c$, we can compute $r_\om$ of Eq.~\eqref{eq:rom} and relate it to $\omm$. 
After a lengthy computation, we obtain
\begin{multline}
\label{eq:rommax}
 r_\om =\frac{3 \sqrt{3}}{2} \,
\frac{\ommax}{\om} \,
\left[ 1 + \frac{D}{6}
- \frac{1}{3\sqrt{3}}\left(1+\frac{D}{2} \right)\frac{\om}{\ommax} 
\right.\\ \left.+ O\left(D^2\right) + O\! 
\left(\frac{\om^2}{\ommax^2}\right)
\right].
\end{multline}
For frequencies $\om/\omm \ll 1$ and small values of $D$, quite remarkably, $r_\om$ depends only on $\ommax/\om$. 

At this point, we remind the reader that, in dispersive theories, it was observed that $\ommax$ governs the first-order deviations from the Hawking spectrum~\cite{MacherRP1,MacherBEC,broad_hor,granada_proc}. 
Since the two negative roots $k^{(i)}_\om$, associated with the negative norm mode, no longer exist for $\om > \omm$, $\ommax$ acts as a cutoff above which there is no radiation.
However, it is more subtle to understand why $\ommax$ also determines the low-frequency ($\om\ll\ommax$) properties of the spectrum.
A detailed analysis~\cite{cpf} of the connection formula governing the scattering on the sonic horizon confirms that the first deviations from the Hawking spectrum are governed by the particular combination of $\Lambda$ and $D$ entering in $\ommax$. From Eq.~\eqref{eq:rommax}, we learn that the same combination fixes the total redshift when $D$ is small. As a result, we shall see that the transition from the Hawking regime to the other one can be meaningfully characterized by $r_\om$ for $\om = \Teff$ of Eq.~\eqref{eq:defT}.

\subsection{Fluxes for steplike profiles}
\label{subsec:step}

When ignoring greybody factors, the spectral properties of $n_\omega$ are analytically known in the Hawking regime which is found in the dispersionless limit ($\Lambda \to \infty$, fixed $\kappa$).
In the opposite limit of infinite surface gravity ($\kappa \to \infty$, fixed $\Lambda$), the spectral properties can be analytically computed~\cite{Recati2009,carlos,Scott_thesis}. In particular, the low-frequency effective temperature $\Teff$ of Eq.~\eqref{eq:defT} is well-defined.
We call it $\Tstep$, and we refer to this case as the {\it steplike regime}. 

Before determining the extension of this domain, we generalize the treatment of the above references so as to work with condensate flows where both $v$ and $c$ have a jump at the sonic horizon.
We focus on the low-frequency behavior of $n_\om$, the spectrum emitted in the subsonic region, {and} $n^v_\om$, the spectrum of left movers with respect to the atoms, see Eq.~\eqref{eq:occnumb}.
In the steplike regime, unlike what is found in the Hawking regime, the spectra of both right and left movers are predicted by the same techniques and should be treated on equal footing.
In fact, whereas the spectrum of left-going particles is not well-defined for gravitational black holes as there is no asymptotic region inside the horizon, it is perfectly well-defined for acoustic black holes. Moreover, the mode mixing of left and right movers has been observed in gravity waves experiments~\cite{Silke} and could be observed in the future in Bose-Einstein condensates.
In Appendices~\ref{app:system} and~\ref{app:step}, we give a summary of the concepts needed to compute these two occupation numbers. 
Here, we present only the main results.

In the low-frequency limit, we obtain
\begin{equation}
 \nstep 
=\frac{\Lambda}{\om}\left(\frac{D}{2}\right)^{3/2}
\left[1-\frac{D}{2}\left(q+\frac{1}{2}\right)+O\left(D^2\right)\right].
\\
\label{eq:n_step}
\end{equation}
Using Eqs.~\eqref{eq:defT} and~\eqref{eq:ommax}, the effective temperature reads
\begin{equation}\label{eq:Tstep}
 \Tstep 
=\frac{3 \sqrt{3}}{8}\,\ommax
\left[ 1 - \left(q-\frac{1}{6}\right)D+ O\left(D^2\right)\right].
\end{equation}
This simple equation establishes the relevance of $\ommax$ in the steplike regime.
In particular, it establishes that $n_\om \propto D^{3/2}$ as $D \to 0$.
This nontrivial result is relevant for experiments in optical fibers and glasses where $D \lesssim 10^{-3}$. 
It is also interesting to express $\nstep$ in terms of $r_\om$ of Eq.~\eqref{eq:rom}:
\begin{equation}
\label{nstep_r}
 \nstep =\frac{r_\om}{4}\left[ 1 - qD+ O\left(D^2\right)\right].
\end{equation}
Hence, for small values of $D$, using Eq.~\eqref{eq:cvu}, we get another simple equation:
\begin{equation}
 \nstep \approx\frac{r_\om \, |\vu|}{4\cn},
\end{equation}
where $\vu$ is the fluid velocity in the upstream region. We notice that $\nstep$ factorizes into $r_\om$ and $\vu$, depending on quantities which are, respectively, defined in the supersonic and subsonic region only.

Operating in a similar way, we obtain the number of left-going particles:
\begin{multline}\label{eq:nv_step}
 \nvstep =\sqrt{2}\left(q-\frac{1}{2}\right)^2\frac{\Lambda}{\om}\,D^{7/2}\\
 \qquad\times\left[1-\frac{D}{2}\left(q+\frac{1}{2}\right)+O\left(D^2\right)\right].
\end{multline}
In analogy with Eq.~\eqref{eq:defT}, we define $\Tveff$ as the effective temperature of left movers.
In the steplike regime, we get
\begin{multline}
 \Tvstep  =\frac{3\sqrt{3}}{2}\, \ommax \,
D^2  \left(q-\frac{1}{2}\right)^2
\left[ 1 - \left(q-\frac{1}{6}\right)D\right]\\
+ O\left(D^4\right).
\label{eq:Tv_step}
\end{multline}
Note that $\Tvstep$ vanishes quadratically when $q=1/2$.
This is reminiscent of what was numerically noticed in Ref.~\cite{MacherBEC} for continuous profiles. In that case, it was found that the mixing between right- and left-going modes was minimum for this value.
For a steplike profile, the first nonvanishing term is
\begin{equation}
 \nvstep 
=\frac{D^{11/2}}{1152\sqrt{2}}\left[1+\frac{D}{2}\right]\frac{\Lambda}{\om}+O\left(D^{13/2}\right).
\end{equation}

Finally, since both $n_\om^v$ and $n_\om$ diverge in the same way for $\om \to 0$, it is convenient to compute their ratio 
\begin{equation}\label{eq:Rdef}
 R_{\rm eff}\equiv\lim_{\om\to0}\left(\frac{n^v_\om}{n_\om}\right).
\end{equation}
For steplike profiles, using Eqs.~\eqref{eq:n_step} and~\eqref{eq:nv_step}, one gets
\begin{equation}\label{eq:Rstep}
 R_{\rm step}=4\left(q-\frac{1}{2}\right)^2 D^2+O\left(D^4\right).
\end{equation}
From this ratio, we learn that $n^v_\om$ is highly suppressed in optical fibers since $D \ll 1$ and that it is not directly related to $r_\om$ of Eq.~\eqref{eq:rom}, unlike what is found for the right movers in Eq.~\eqref{nstep_r}. In that case indeed, for $D \ll 1$, both $\Tstep$ and $r_\om$ have the same dependence on $\Lambda$, $D$, and $q$.

In the next section, we determine the validity domain of these formulas. 
It is rather clear that the above expressions should be valid in the limit of infinite surface gravity: $\kappa/\Lambda \to \infty$ at fixed $D$. 
Rather surprisingly, we shall see that they are also valid for very small $\kappa/\Lambda$ when $D\to 0$.

\section{The extension of the two regimes}
\label{sec:numerics}

\subsection{The Hawking regime}

%
\begin{figure}[b]
 \includegraphics[width=0.85\columnwidth]{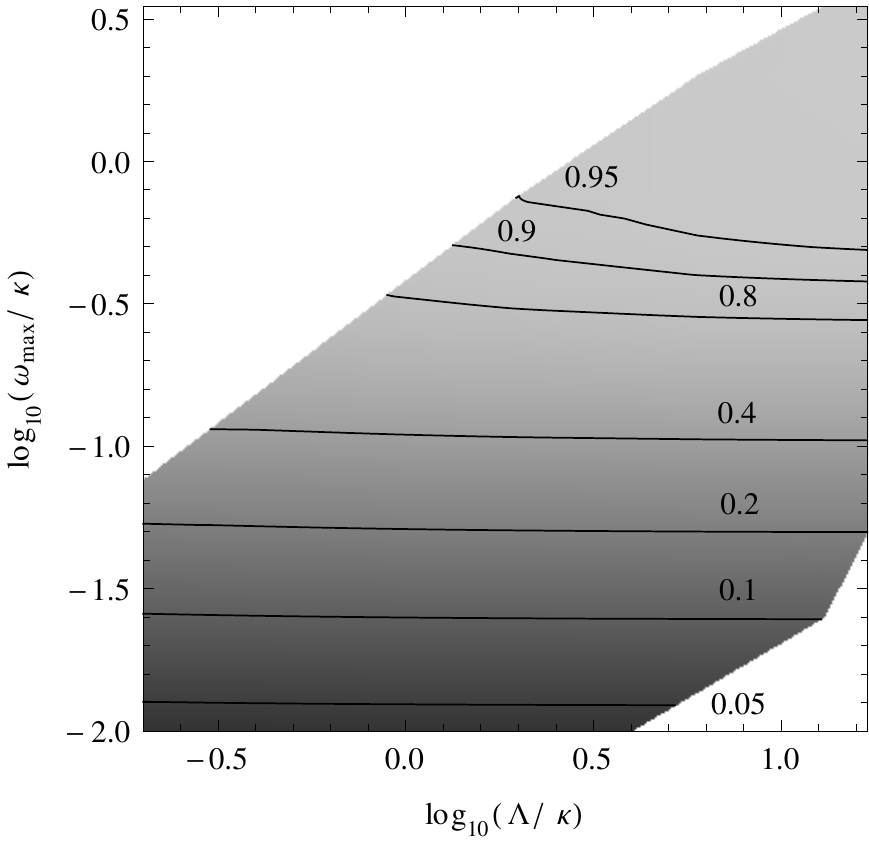}
 \caption{$\Teff/\Th$ as a function of $\Lambda/\kappa$ and $\ommax/\ka$ in logarithmic scales. The parameter $q$ of Eq.~\eqref{eq:cv} is $0.3$. Solid lines are loci of constant $\Teff/\Th$.
One clearly sees that the departures from the Hawking regime are essentially governed by $\omm/\kappa$. 
Roughly speaking, one leaves the Hawking regime (i.e., $\Teff$ and $\Th$ differ by 10\%)
for $\ommax/\kappa \lesssim 0.4$, 
irrespectively of the value of $\Lambda/\kappa$. Note that
the upper white area is due to the fact that we restrict our analysis to $D<1$, whereas the lower one is due to our incapacity of numerically obtaining $n_\om$ for $D \lesssim 0.02 $.}
 \label{fig:TeffoTh}
\end{figure}
\begin{figure*}[t]
\includegraphics[width=0.85\columnwidth]{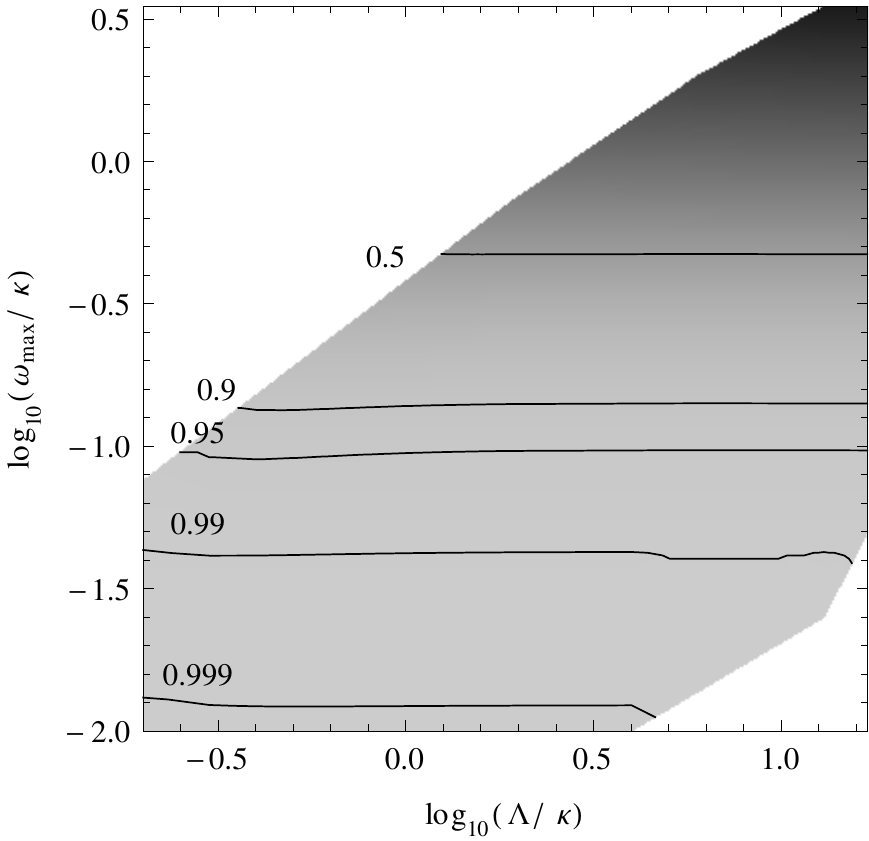}
\hspace{2.em}
 \includegraphics[width=0.85\columnwidth]{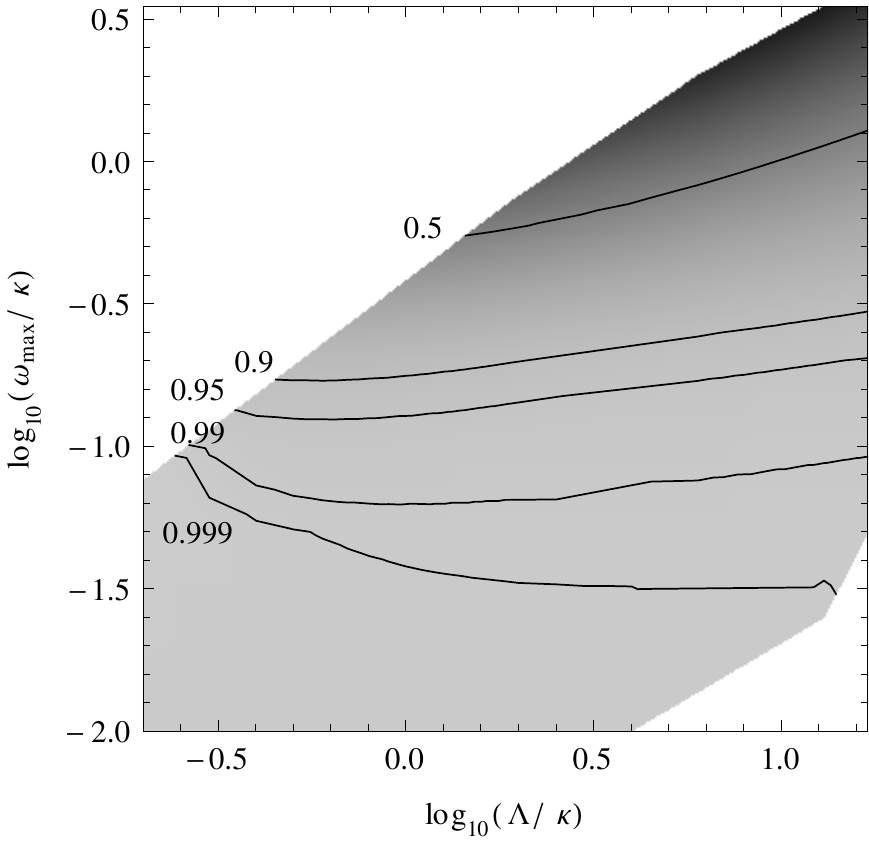}
 \caption{The temperature ratios $\Teff/\Tstep$ (left panel) and $T^v_{\rm eff}/\Tstep^v$ (right panel) 
as a function of $\Lambda/\kappa$ and $\ommax/\ka$ for the same situations as in Fig.~\ref{fig:TeffoTh}, where $\Tstep$ and $\Tstep^v$ are obtained from Eqs.~\eqref{eq:beta} and \eqref{eq:B}.
The departures from the steplike regime are essentially governed by $\omm/\kappa$. Roughly speaking, one leaves this regime for $\ommax/\kappa \gtrsim 0.1$, irrespectively of the value of $\Lambda/\kappa$.}
\label{fig:TeffoTstep}
\end{figure*}
We start with the extension of the Hawking regime. This has been already studied, and it is rather well-understood. 
Using numerical techniques~\cite{MacherBEC}, it was found that the first-order deviations from the Hawking spectrum are mainly controlled by $\ommax/\kappa$. They thus scale as $\Lambda D^{3/2}$, see Eq.~\eqref{eq:ommax}.
In fact, it was found that these deviations hardly depend on $\Lambda/\kappa$ when working at fixed $\ommax/\kappa$. 
Using the same code, we now clarify these observations and also investigate the spectral properties further away from the Hawking regime.
To these ends, we plot in Fig.~\ref{fig:TeffoTh} the ratio $\Teff/\Th$ in the two-dimensional parameter space defined by $\ommax/\kappa$ and $\Lambda/\kappa$. We use logarithmic scales to see more easily the scaling properties. The value of the parameter $q$ has been chosen to be 0.3, to avoid the peculiar cases of $q=1/2$ or $q=1/6$, where some contribution cancels, as can be seen from the expressions given in Sec.~\ref{subsec:step}.

For $\ommok \gtrsim 1$, we see that $\Teff/\Th \to 1$ in a manner that hardly depends on $\Lambda/\kappa$. The region  $\ommok\gtrsim 1$ therefore characterizes the situations where Hawking's result is robust.
For $\ommok \ll 1$, the effective temperature $\Teff$ strongly differs from $\Th$. However, it still hardly depends on the dispersive scale $\Lambda/\kappa$.
Therefore, $\ommok$ is the most relevant spectral parameter in the {\it entire plane}.
This important lesson is corroborated by the results presented in the following subsection.

\subsection{The steplike regime}

In a manner strictly similar to what we have just done for the Hawking regime, we study the validity domain of the expressions obtained in Sec.~\ref{subsec:step}.
In Fig.~\ref{fig:TeffoTstep}, we plot $\Teff/\Tstep$ (left panel) and $\Tveff/\Tvstep$ (right panel) for the same situations as in Fig.~\ref{fig:TeffoTh}. 
The first lesson from Fig.~\ref{fig:TeffoTstep} is that these two ratios behave in a very similar manner.
Second, when $\ommok \lesssim 0.1$, we see that the effective temperatures $\Teff$ and $\Tveff$ obtained with the regular profiles of Eq.~\eqref{eq:velocity} are in excellent agreement with $\Tstep$ and $\Tvstep$ obtained in Sec.~\ref{subsec:step}. 
In addition, we learn that this agreement is reached in a manner that hardly depends on $\Lambda/\kappa$. 
Therefore, quite remarkably, the steplike regime is found even when scale separation is achieved,
i.e., for $\Lambda/\kappa \gg 1$, when $D$ is small enough that $\ommax/\kappa \lesssim 0.1$.

\begin{figure}[b]
 \includegraphics[width=0.85\columnwidth]{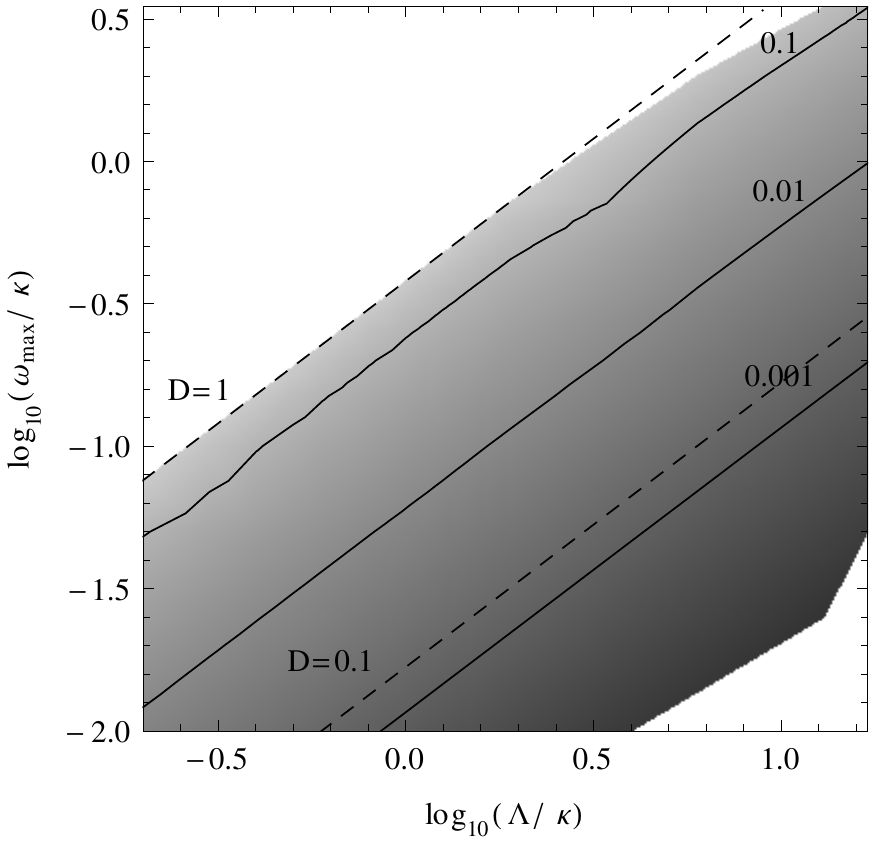}
 \caption{$R_{\rm eff}$ defined in Eq.~\eqref{eq:Rdef} as a function of $\Lambda/\ka$ and $\ommax/\ka$ 
for the same situations as in Fig.~\ref{fig:TeffoTh}.
Solid lines give the various values of this ratio, whereas the dashed straight lines with slope $2/3$ are loci of constant $D$, see Eq.~\eqref{eq:ommax}. 
Clearly, $R_{\rm eff}$ depends only on $D$.}
 \label{fig:nonv}
\end{figure}
To complete our analysis, in Fig.~\ref{fig:nonv}, we plot the ratio $R_{\rm eff}$ of Eq.~\eqref{eq:Rdef}.
We note that the lines of constant $R_{\rm eff}$ almost coincide with the lines of constant $D$ (slope $2/3$, dashed lines).
Hence, $R_{\rm eff}$ scales as $D^2$.
This scaling is in perfect agreement with the steplike prediction of Eq.~\eqref{eq:Rstep}.
This is further confirmation of the applicability of steplike techniques in a wide region of parameters.
However, unlike $\Teff$ and $\Tveff$ which are well-approximated by using a steplike profile only for $\ommax/\kappa \lesssim 0.1$, their ratio behaves as $D^2$ in the whole plane. This means that $R_{\rm eff}$ scales in the same way in the Hawking and in the steplike regimes.

We also emphasize that  $R_{\rm eff}$ is maximal for large values of $D$. 
This is also rather unexpected because the coupling between right and left sectors is a nonadiabatic effect. 
The coupling should thus be smaller for smoother profiles~\cite{CJ96}.
When adding high-frequency dispersion to a two-dimensional conformally invariant massless field~\cite{Unruh95,BMPS95}, this is indeed the case, see Fig. 19 in Ref.~\cite{MacherRP1}. 
However, when using the quasiparticle wave equation of some condensed matter model, 
the opposite behavior is found, see e.g. Figs.~11 and~12 of \cite{MacherBEC}.
Combining these results with the present analysis of step functions, it is now clear that the spectrum $n^v_\om$ of left movers increases with $D$ even though the profile is smoother for larger $D$.

\section{Summary and interpretation}

\begin{figure}[b]
 \includegraphics[width=0.85\columnwidth]{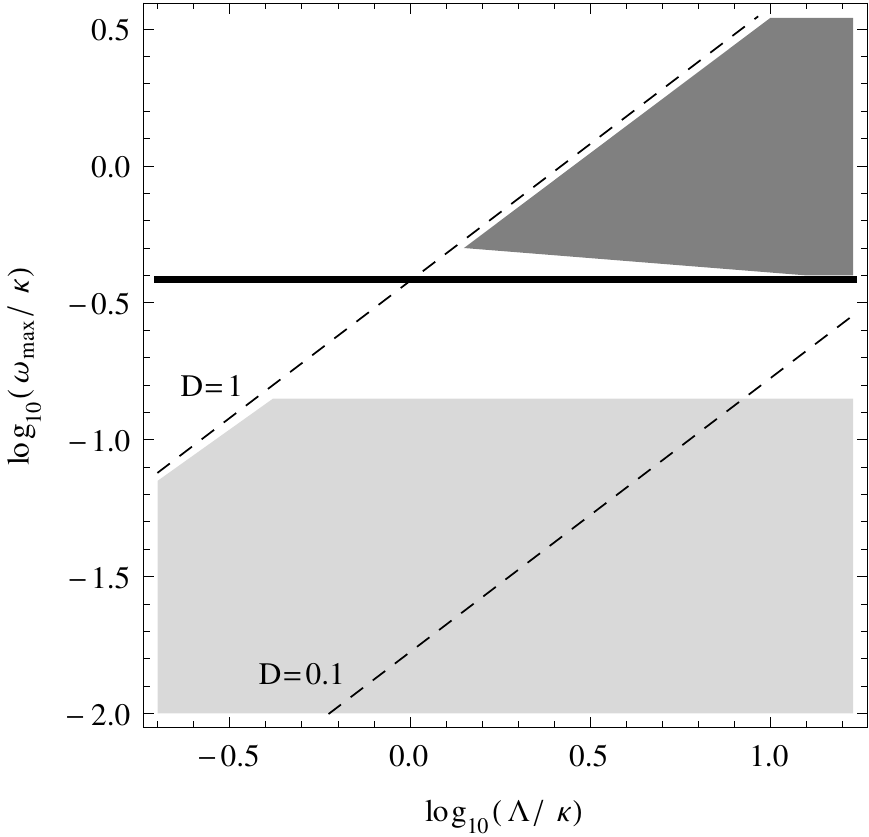}
 \caption{In dark gray, we denote the validity region (defined by 10\% difference in temperature) of the Hawking regime, and in light gray, that of the steplike regime.
The thick horizontal line is the threshold value of Eq.~\eqref{eq:thres}.
The dashed straight lines are loci of constant $D$. It is clear that $\omm/\kappa$ governs the extension of the two domains and the transition between them.}
 \label{fig:D}
\end{figure}

To summarize the results, we schematically illustrate in Fig.~\ref{fig:D} the validity regions of the two regimes. 
The region where the temperature $\Teff$ differs from $\Th$ by less than 10\% is shaded in dark gray, whereas that approximately governed by a steplike profile is shaded in light gray.
We reemphasize that the expressions obtained using a steplike profile provide very good approximations not only where $\ka/\Lambda$ is large (right lower corner), but in general when $\ommax/\ka$ is small, independently of the value of $\ka/\Lambda$. 

In Fig.~\ref{fig:D}, we see that the white area from $0.1\lesssim\ommok\lesssim0.4$ is not covered by either approximation. To characterize the effective temperature in this transitory regime, we use the {\it average} temperature $\Tav$ of Ref.~\cite{broad_hor}. It is obtained by taking the spatial average of the gradient of $w = c + v$ over a given length $d_\xi$,
\begin{equation}\label{eq:Tav}
 \Tav\equiv\int_{-d_\xi/2}^{d_\xi/2}dx\, \frac{dw}{dx}=\frac{2w(d_\xi/2)}{d_\xi}.
\end{equation}
The critical length was found to be
\begin{equation}\label{eq:dxi}
 d_\xi=d\,\chor \left(\kappa\Lambda^2\right)^{-1/3},
\end{equation}
where $d$ is approximately equal to $2^{2/3}$. 
When the extension of the NHR where $w \sim \kappa x$ is larger than $d_\xi$, $\Tav$ agrees with the Hawking temperature $\Th$. 

Transposing these notions to the present context, the boundary of the region where the Hawking regime ceases to be valid should thus be given by the condition that the extension of the NHR of Eq.~\eqref{eq:velocity} equals $d_\xi$: 
\begin{equation}\label{eq:resultbroad}
 2\frac{D\chor}{\ka} = d_\xi. 
\end{equation}
Using Eqs.~\eqref{eq:dxi} and~\eqref{eq:ommax}, this condition fixes the threshold value of $\omm/\ka$ to be 
\begin{equation}\label{eq:thres}
 \frac{\ommax}{\ka}|_{\rm threshold} = \frac{2}{3\sqrt{3}}.
\end{equation}
In Fig.~\ref{fig:D}, the black solid line represents this value. 
In accord with the analysis of~\cite{broad_hor,cpf}, it precisely lies where the Hawking regime ceases to be valid.

\begin{figure}[b]
\includegraphics[width=0.85\columnwidth]{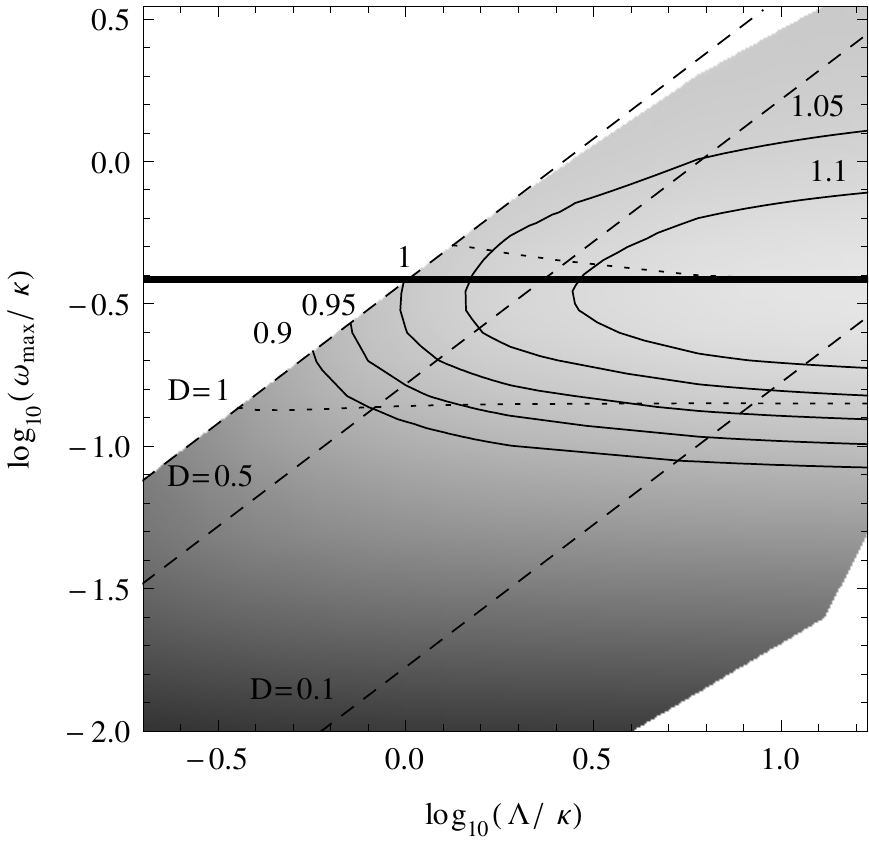}
\caption{$\Teff/\Tav$ as a function of $\Lambda/\kappa$ and $\ommax/\ka$ in logarithmic scale for the same profiles as in Fig.~\ref{fig:TeffoTh}, where $\Tav$ is given by Eqs.~\eqref{eq:Tav} and \eqref{eq:dxi}.
Solid lines are loci of constant $\Teff/\Tav$. 
The other lines are as in Fig.~\ref{fig:D}.}
\label{fig:TeffoTav}
\end{figure}
To be more precise and more quantitative, we now compare $\Tav$ with the effective temperature $\Teff$. Their ratio is plotted in Fig.~\ref{fig:TeffoTav}. As expected, $\Tav$ is very close to $\Teff$ when $D$ and $\Lambda/\kappa$ are large. 
When $D \ll 1$, $\Tav$ is a bit less accurate in reproducing $\Teff$ than $\Th$, even if $\ommax/\kappa$ is large. 
However, $\Tav$ gives a rather good approximation of $\Teff$ in a much broader region than $\Th$.
Within an error of 10\%, the region covered by $\Tav$ extends down to $\log_{10}\ommax/\kappa\approx-1$.
The average temperature $\Tav$ therefore correctly describes $\Teff$ in the transition region where both $\Th$ and $\Tstep$ fail to do so.

\begin{figure}[t]
 \includegraphics[width=0.85\columnwidth]{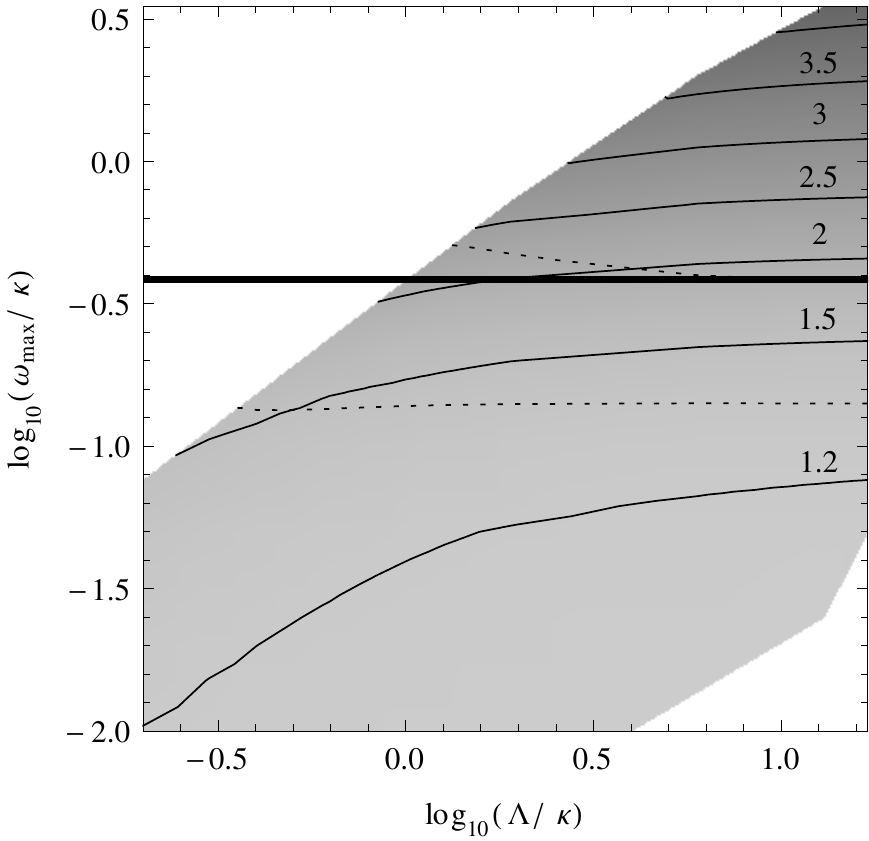}
 \caption{The number of $e$-folds of Eq.~\eqref{eq:rom} evaluated for $\om = \Teff$ as a function of $\Lambda/\kappa$ and $\ommax/\ka$ for the same profiles as in Fig.~\ref{fig:TeffoTh}.
The dotted lines bound the regions of validity (with 10\% errors) of the Hawking regime (upper line) and the steplike regime (lower line). The thick line is the threshold value of Eq.~\eqref{eq:thres}.
The Hawking and the steplike regime are, respectively, found for $N_{\Teff} \gtrsim 2$ and 
 $N_{\Teff} \lesssim 1.3$. }
 \label{fig:Nom}
\end{figure}

Now, what is the physical meaning of Eq.~\eqref{eq:thres}?
Using Eq.~\eqref{eq:rommax}, we see that $\ommax$ is proportional to $\om\,r_\om$.
To be able to define a threshold value for the number of $e$-folds $N_\om = \ln r_\om$, we need to pick a frequency $\om$. In the sequel, we work with $\om = \Teff$ since $\Teff$ characterizes the actual spectrum $n_\om$ for $\om \to 0$.
When $\ommax < \ommax|_{\rm threshold}$, we see in Fig.~\ref{fig:Nom} that $N_{\om = \Teff}$ is almost constant and slightly larger than 1. It has to be noticed that it does {\it not} go to $0$ as $D \to 0$ when working with $\om = \Teff$.
When $\omm$ is above the threshold, $N_{\Teff}$ increases proportionally to 
$\log_{10}(\omm/\ka)$.
We therefore conclude that the number of $e$-folds evaluated at the typical frequency $\Teff$ provides an extremely simple way to characterize the validity domains of the two regimes.
That is, the Hawking regime starts above $N_{\Teff} \sim 2$, whereas the steplike analysis provides reliable results when $N_{\Teff}$ is smaller than $1.4$.

To conclude this paper, we first underline that our results, which have been obtained using the superluminal quartic dispersion 
of Eq.~\eqref{eq:dispersion_hom}, also apply to other types of superluminal relations, and to subluminal ones, in virtue of the correspondence between these cases~\cite{cpf}.

Second, we point out that our analysis can be used to characterize the outcome of recent and future experimental investigations aiming to detect the analogue Hawking effect.
Indeed, the knowledge of $\Lambda/\kappa$ and $\ommax/\kappa$ suffices to determine in which regime one will find the (low-frequency part of the) quasiparticle spectrum.
For instance, a steplike analysis should be appropriate for experiments in optical fibers or glasses since these are characterized by a small parameter $D$, typically of order $\lesssim 10^{-3}$~\cite{ulf,faccio}, in spite of the fact that $\Lambda/\kappa \gg 1$.
Thus, in these cases, a precise knowledge of the surface gravity is not required to predict the spectrum.
On the contrary, for experiments using as analogue system phonons in Bose-Einstein condensates~\cite{jeff} or gravity waves on water~\cite{Silke}, the spectrum will be essentially governed by the surface gravity because $D$ is of the order of unity, and $\Lambda/\kappa > 1$.
Unfortunately, a more precise estimate of these parameters cannot be provided for the later experiment
because the background profile of the free surface is not well-understood. The main reason is that it 
contains a macroscopic undulation, a zero-frequency wave related to the analogue
Hawking effect in white hole flows~\cite{Carusottowhite,cpf}.

Third, our analysis can be used to characterize the Hawking radiation in theories of quantum gravity~\cite{Ted_EinstAether,Parentani2001,HoravaPRD79,Blas_Sibiryakov2011} which contain violations of Lorentz invariance at very high energies.
In particular, Fig.~\ref{fig:TeffoTh} indicates that the black hole temperature should no longer be given by the surface gravity when $\kappa/\Lambda$ becomes larger than 1.
Understanding the possible consequences of this on black hole thermodynamics~\cite{Ted_wall_review} is an interesting and challenging question.

\begin{acknowledgments} 

S.~F. thanks I. Carusotto for helpful suggestions and A. Recati for explanations about Ref.~\cite{Recati2009}.
R.~P. wishes to thank U. Leonhardt for conversations (in Winter 2004) on Hawking radiation in optical fibers during which he realized that the parameter $D$ and the number of $e$-folds must play a crucial role.
We both thank A. Coutant for a careful reading of this work.
S.~F. has been supported by the Foundational Questions Institute (FQXi) Grant No.~FQXi-MGA-1002.

\end{acknowledgments}

\appendix

\section{Mode analysis and scattering matrix}
\label{app:system}

We present the main concepts which characterize the scattering of phonons in atomic Bose condensates in stationary flows containing one sonic horizon. We follow the treatment of Ref.~\cite{MacherBEC} where more details can be found.

In nonhomogeneous condensates, phonon elementary excitations are appropriately described by {\it relative} density perturbations:
\begin{equation}
 \hP=\Pn(1+\hp),
\end{equation}
where $\hP$ is the atom field operator, and  $\Pn$ the condensate.
The dynamics of $\hp$ is determined by the Bogoliubov-de~Gennes equation~\cite{DalfovoRMP}:
\begin{equation}\label{eq:bdg}
  \ii\pdt\hp = \left[T_\rho-\ii v \pdx + m c^2 \right] \hp + m c^2 \hpd,
\end{equation}
where $\hbar = 1$, $c$ is the $x$-dependent speed of sound
\begin{equation}
 c^2(x)\equiv\frac{g(x)\rn(x)}{m},
\end{equation}
$\rn(x)= |\Pn(x) |^2$ is the mean density of condensed atoms, and $g(x)$ is the effective coupling constant among atoms. These functions play no role in the sequel: only $c(x)$ and $v(x)$ matter. The kinetic operator that acts on $\hp$ is
\begin{equation}\label{eq:Trho}
 T_\rho=-\frac{ 1 }{2m} \, v\pdx \frac{1}{v}\pdx.
\end{equation}

In stationary situations, $\hp$ can be expanded in frequency eigenmodes
\begin{multline}\label{eq:phiexpansion}
\hp =  \int\!\dom\left[\phm{\om t}\hp_\om(x)+\php{\om t} \hat \varphi_\om(x)^\dagger\right] \\ 
=\int\!\dom\sum_{\alpha}\left[\phm{\om t}\pom\aom+\php{\om t}\vpoms\aomd\right],
\end{multline}
where the discrete sum over $\alpha$ takes into account the number of modes at fixed $\om$.
Inserting Eq.~\eqref{eq:phiexpansion} in Eq.~\eqref{eq:bdg} yields the following system:
\begin{equation}\label{eq:system}
\begin{aligned}
 \left[ (\om +\ii v\pdx) - T_\rho - mc^2\right]\pom=  mc^2\vpom,\\
 \left[- (\om +\ii v\pdx) - T_\rho - mc^2\right]\vpom = mc^2\pom.
 \end{aligned}
\end{equation}

In backgrounds containing one sonic horizon, there are two or three modes depending on whether $\om$ is larger or smaller than $\omm$ of Eq.~\eqref{eq:ommax}.
More precisely, the $\om$ component of the field operator reads
\begin{equation}\label{eq:planewaves}
 \hp_\om =\phi_\om^u \ha_\om^u+\phi_\om^v\ha_\om^v + \theta(\ommax - \om) \varphi_{-\om}^*\ha_{-\om}^{\dagger}.
\end{equation}
\begin{figure}
 \includegraphics[width=\columnwidth]{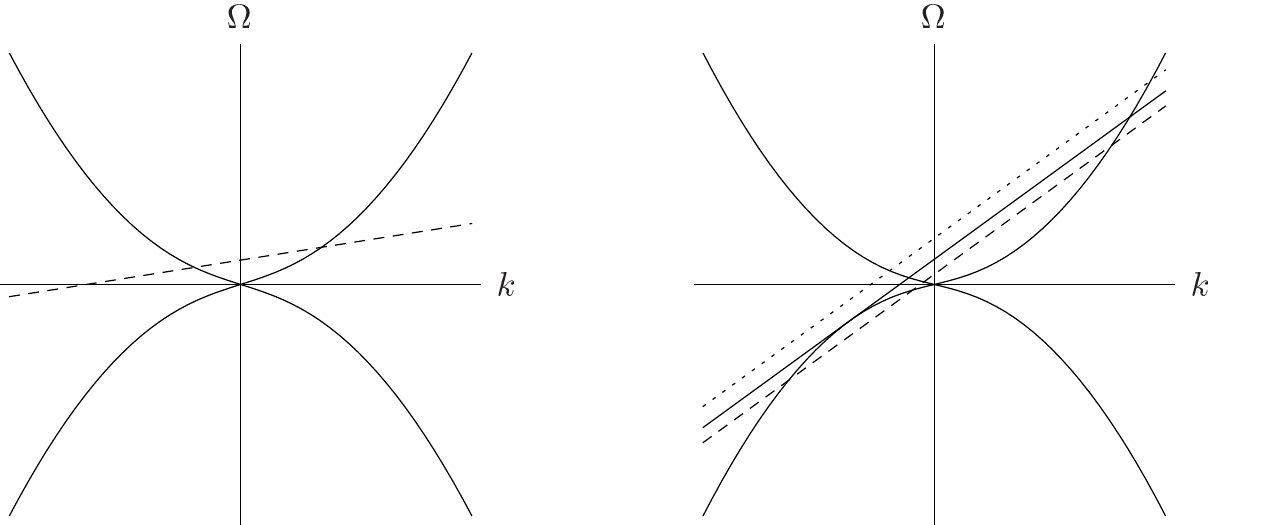}
 \caption{Graphical solution of the dispersion relation~\eqref{eq:dispersion_hom} for subsonic (left panel) and supersonic flow (right panel). Solid curves: $\pm \Omega(k)$. In subsonic flows, for $\om > 0$, the dashed line $(\om-vk)$ crosses twice the dispersion relation with $\Omega > 0$.
In supersonic flows, two extra roots with $\Omega<0$ exist in the left lower quadrant for $\om<\omm$ (dashed line).
When $v$ and $c$ take their asymptotic values in the limit $x\to-\infty$, the most (least) negative of these two roots corresponds to the wave number $\kone$ ($\ktwo$) of the ingoing (outgoing) wave, appearing in Eq.~\eqref{eq:rom}.
For $\om=\omm$, the solid line is tangent to $\Omega(k)$, and for $\om > \omm$ (dotted line), these real roots no longer exist.}
 \label{fig:dispersion}
\end{figure}
For $\om > 0$, the modes $\phi_\om^u$ and $\phi_\om^v$ have positive norm, they describe, respectively, propagating right-$u$ and left-$v$ moving waves with respect to the condensed atoms, and they are associated with the usual real roots of Eq.~\eqref{eq:dispersion_hom}, see Fig.~\ref{fig:dispersion}.
On the contrary, $(\varphi_{-\om})^*$ has a negative norm.
It is associated with the extra roots which exist in supersonic flows for $\om < \omm$ and describes phonons which are trapped inside the horizon ($x<0$) in the supersonic region.

In our infinite condensates, the above modes become superpositions of plane waves both in the left and right asymptotic regions of Eq.~\eqref{eq:planewaves}.
We can thus construct without ambiguity the in- and out-mode bases.
The in (out) modes are such that each of them contains only one asymptotic branch with group velocity directed towards (away from) the horizon. 
For $\om>\omm$, there exists only two positive norm modes, and therefore in and out modes are linearly related by a trivial (elastic) $2 \times 2$ transformation.
Instead, when  $\om < \omm$, the in and out modes mix with each other in a nontrivial way by a $3 \times 3 $ Bogoliubov transformation~\cite{MacherBEC}:
\begin{equation}\label{eq:bog_transf}
\begin{aligned}
 \phi_\om^{u,{\rm in}} &= \alpha_\om \phi_\om^{u,\rm out} + \beta_{-\om} \left(\varphi_{-\om}^{\rm out}\right)^*
 				+ \tilde A_\om \phi_\om^{v,\rm out},\\
 (\varphi_{-\om}^{{\rm in}})^* &= \beta_\om \phi_\om^{u,\rm out} + \alpha_{-\om} \left(\varphi_{-\om}^{\rm out}\right)^* 
 				+ \tilde B_\om \phi_\om^{v,\rm out} ,\\
 \phi_\om^{v,{\rm in}} &= A_\om \phi_\om^{u,\rm out} + B_{\om} \left(\varphi_{-\om}^{\rm out}\right)^*
 				+ \alpha_\om^v \phi_\om^{v,\rm out}.
\end{aligned}
\end{equation}
The standard mode normalization yields relations such as (from the first equation)
\begin{equation}
 |\alpha_\om|^2 - |\beta_{-\om}|^2 + |\tilde A_\om|^2 = 1.
\end{equation}
When the initial state is vacuum, the final mean occupation numbers are given by 
\begin{equation}\label{eq:occnumb}
 n_\om = |\beta_\om|^2,\quad
  n_\om^{v} = |\tilde B_\om|^2,\quad
  n_{-\om} =  n_\om +  n_\om^{v}.
\end{equation}
The third equation, which gives the number of negative frequency phonons,
follows from energy conservation which imposes that each pair spontaneously produced should have no energy.
In the standard analysis without dispersion, $n_\om$ is Planckian and describes the Hawking effect~\cite{Primer}.
In that case, for massless conformally invariant fields, the number $n_\om^{v}$ of left-moving phonons identically vanishes because the coupling between $u$ and $v$ modes is zero.
Adding high-frequency dispersion to that case, one finds $ n_\om^{v} \ll n_\om$. 
Hence, the scattering of this dispersive field on a sonic horizon basically consists of a more general and slightly modified version of the standard Hawking effect, as long as $\omm \gtrsim \ka$.
For larger values of $\ka$, and for nonconformally invariant fields, deviations with respect to the relativistic case can be large, but the structure of Eq.~\eqref{eq:bog_transf} and the meaning of its coefficients remain unchanged~\cite{MacherBEC,Recati2009,carlos}.

\section{Steplike profiles}
\label{app:step}

We extend the computation of the Bogoliubov coefficients given in~\cite{Recati2009,carlos} to steplike profiles where the velocity is different on the upstream and downstream sides of the horizon. That is, we consider profiles given by
\begin{align}
 v(x)&=\vu\,\thp+\vd\,\thm,\\
 c(x)&=\cu\,\thp+\cd\,\thm.
\end{align}
For these profiles, Eq.~\eqref{eq:system} can be solved separately in the upstream ($x>0$) and downstream ($x<0$) regions by Fourier transform:
\begin{align}
 \phi_i(x)&=\sqrt{\frac{\partial k_i}{\partial\om}}\,\frac{\ee^{-\ii\om t+\ii k_i x}}{\sqrt{4\pi\rho_i}}\,u_i,\\
 \varphi_i(x)&=\sqrt{\frac{\partial k_i}{\partial\om}}\,\frac{\ee^{-\ii\om t+\ii k_i x}}{\sqrt{4\pi\rho_i}}\,v_i,
\end{align}
where the index $i$ stands for $u$ (upstream) when $x>0$ and $d$ (downstream) when $x<0$. 
(For the sake of conciseness, we have omitted the indices $\alpha$ and $\omega$).
The normalization factor has been chosen such that
\begin{equation}
 u_i^2-v_i^2=1,
\end{equation}
where $u_i,v_i$ are chosen real.
On each side, Eq.~\eqref{eq:system} gives
\begin{equation}\label{eq:systemuv}
\begin{aligned}
 \left[ (\om -v_i\,k_i) - \frac{k_i^2}{2m} - mc_i^2\right]u_i=  mc^2\,v_i,\\
 \left[- (\om -v_i\,k_i) -\frac{k_i^2}{2m}  - mc_i^2\right]u_i = mc^2\,u_i.
 \end{aligned}
\end{equation}
Since the dispersion relation is quartic, we obtain four roots $k(\om)$.
Hence, for a given frequency $\om$, the general solution can be written on either side as superposition of four momentum eigenmodes. Notice that $k$ are all real in the downstream supersonic region, whereas two are real and the other two complex conjugated in the upstream region.

In order to obtain globally defined solutions, we need to match the two sets of modes.
First, the solutions $\phi_{\om,i}(x)$ and $\varphi_{\om,i}(x)$ must be continuous at the horizon, that is
\begin{equation}\label{eq:match}
 \phi_{u}^+=\phi_d^-,\quad\varphi_u^+=\varphi_d^-,
\end{equation}
where the superscript $\pm$ denotes the limit of the function for $x\to0^\pm$, respectively.
Second, the matching conditions for the derivatives of $\phi$ and $\varphi$ can be obtained by integrating the system~\eqref{eq:system} in a small neighborhood $[-\varepsilon,\varepsilon]$ of the horizon and taking the limit for $\epsilon\to0$. 
To characterize this limit, it is useful to write $T_\rho$ of Eq.~\eqref{eq:Trho} as
\begin{equation}
 T_\rho=-\frac{ 1 }{2m}\pdx^2 +\frac{ 1 }{2m} \left[\pdx\log(v(x))\right]\pdx.
\end{equation}
For our profiles, the second term gives
\begin{equation}
\begin{aligned}
 \pdx\log(v(x))&=\pdx\left[\log(\vu)\,\thp+\log(\vd)\,\thm\right]\\
 &=\log(\vu)\,\delta(x)-\log(\vd)\,\delta(-x)\\
 &=\log\!\left(\frac\vu\vd\right)\delta(x).
\end{aligned}
\end{equation}
The nontrivial terms of the first equation of Eq.~\eqref{eq:system} are 
\begin{multline}
 \ii\int_{-\varepsilon}^\varepsilon\dx\, v\,\partial_x\phi
+\frac{1}{2m}\int_{-\varepsilon}^\varepsilon\dx\,\partial_x^2\phi\\
-\frac{1}{2m}\log\!\left(\frac\vu\vd\right)\int_{-\varepsilon}^\varepsilon\dx\,\delta(x)\,\partial_x\phi=0.
\end{multline}
Since $\phi$ is continuous at $x=0$, its derivative can have at most a finite jump at $x=0$. Therefore, the integrand in the first integral has a finite jump at $x=0$, such that, when taking the limit $\varepsilon\to0$, this term vanishes. The second integral can be directly computed because its integrand is the derivative of $\partial_x\phi$. It gives the jump in the first derivative of the field $\phi$:
\begin{equation}
 \int_{-\varepsilon}^\varepsilon\dx\,\partial_x^2\phi=\partial_x\phi_u^+-\partial_x\phi_d^-.
\end{equation}
Finally, the last integral gives
\begin{equation}
 \log\!\left(\frac\vu\vd\right)\int_{-\varepsilon}^\varepsilon\dx\,\delta(x)\,\partial_x\phi
 = \frac{1}{2}\log\!\left(\frac\vu\vd\right)\left[\partial_x\phi_u^++\partial_x\phi_d^-\right].
\end{equation}
Taking into account these last equations, the matching condition of the first derivative is
\begin{equation}
 \left[1-\frac{1}{2}\log\!\left(\frac\vu\vd\right)\right]\partial_x\phi_u^+
 =\left[1+\frac{1}{2}\log\!\left(\frac\vu\vd\right)\right]\partial_x\phi_d^-.
\end{equation}
From the second equation of~\eqref{eq:system}, one finds an identical condition for $\partial_x\varphi$.
In brief, we get
\begin{equation}\label{eq:matchder}
 \phi_u^+=\eta\,\phi_d^-,\quad \varphi_u^+=\eta\,\varphi_d^-,
\end{equation}
where
\begin{equation}
 \eta\equiv\frac{2-\log(\vd/\vu)}{2-\log(\vu/\vd)}.
\end{equation}
Using Eqs.~\eqref{eq:match} and~\eqref{eq:matchder}, for $\om < \ommax$, one can determine the three globally defined solutions which are {\it asymptotically bounded} and which appear in Eq.~\eqref{eq:planewaves}.
From them, we can extract the nine coefficients of Eq.~\eqref{eq:bog_transf}. The computation is rather tedious, but completely analogous to what was performed in Ref.~\cite{Recati2009,carlos}.
Below we only give the dominant term in the limit $\om\to0$ of two Bogoliubov coefficients $\beta_\om$ and $\tilde B_\om$ which determine $n_\om$ and $n^v_\om$. We obtain
\begin{gather}
\begin{aligned}
\beta_\om&=\sqrt{\frac{\Lambda}{\omega}}
\times
\frac{\sqrt{\cu}}{\chor}\,\sqrt{\frac{\vd}{\vu}}\,
\sqrt{\frac{\cu+\vu}{\cu-\vu}}\,
\\
&\quad\times
\frac{\left(\vd^2-\cd^2\right)^{3/4}}{\cu^2-\cd^2+\cu(\vu-\vd/\eta)}
\\
&\quad\times
\left[\sqrt{\cu^2-\vu^2}-\ii\,\frac{\vd\vu-\eta\,\cd^2}{\sqrt{\vd^2-\cd^2}}\right],
\end{aligned}
\label{eq:beta}\\
\begin{aligned}
\tilde B_\om&=\sqrt\frac{\Lambda}{\omega}\times
\frac{1}{2\chor\sqrt{\cd}}\,
\sqrt{\frac{\cu+\vu}{\cu-\vu}}\,
\\
&\quad\times
\frac{\left(\vd^2-\cd^2\right)^{3/4}}{\left[\cu^2 -\cd^2 +\cu  (\vu-\vd/\eta)\right]}
\\
&\quad\times
\left[\sqrt{\cu^2-\vu^2} \left(\cu -\cd \frac{\cu/\eta+ \vd}{\cu+\vu}\right)
\right.\\
&\quad\left.
-\ii\,  \sqrt{\vd^2-\cd^2}  \left(\cu \frac{\cd \eta+\vu}{\cd+\vd}-\cd \right)\right].
\end{aligned}
\label{eq:B}
\end{gather}

Using our parameterization of the velocity profile given in Eqs.~\eqref{eq:cv} and~\eqref{eq:velocity}
\begin{align}
 \cu &=\cn\left[1+(1-q)D\right],\quad&\vu &=-\cn(1-q\,D), \label{eq:cvu} \\
 \cd &=\cn\left[1-(1-q)D\right],\quad&\vd &=-\cn(1+q\,D),
\end{align}
in the limit of small $D$, we obtain Eqs.~\eqref{eq:n_step} and~\eqref{eq:nv_step}.

\section{Asymmetric profiles}
\label{app:D2}

In this appendix, we show that the results obtained in the main body of the text are still valid when considering flow profiles that generalize Eq.~\eqref{eq:velocity}:
\begin{equation}\label{eq:velocity_as}
 \frac{w_\alpha(x)}{\chor} =  D\left[\alpha+
\tanh \left(\frac{\ka x}{\chor D}\right)\right].
\end{equation}
The parameter $\alpha$ fixes the asymmetry between the sub- and supersonic flows with respect to the sonic horizon.
This horizon exists only if $|\alpha|<1$, and it is located at
\begin{equation}
 x_{\rm H}=-\frac{Dc_0}{\ka}\,\mbox{arctanh}(\alpha).
\end{equation}
The Hawking temperature corresponding to the velocity profile of Eq.~\eqref{eq:velocity_as} is $\Th={\kk}/{2\pi}$, where
\begin{equation}\label{eq:temprel}
\kk  \equiv \partial_x w_\alpha \vert_{x = x_H} = \ka
\left[1-\alpha^{2}\right].
\end{equation}

\begin{figure}[b]
\includegraphics[width=0.85\columnwidth]{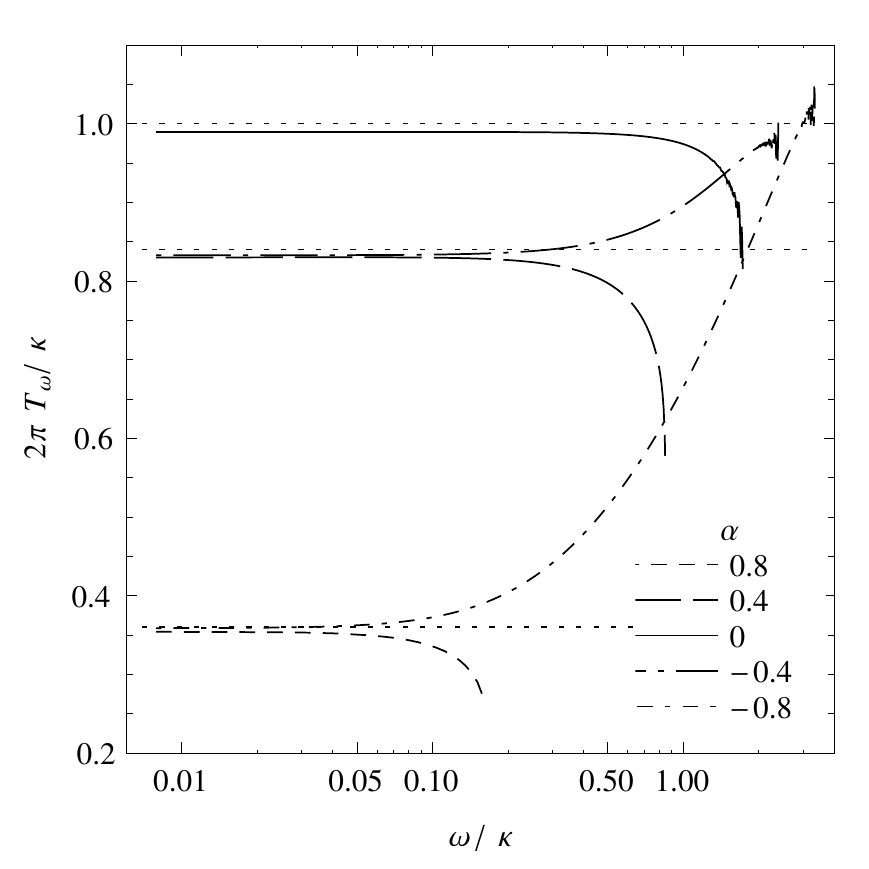}
\caption{\label{fig:tom}$T_\om$ for 5 values of $\alpha$, with $D=0.5$ and $\Lambda/\kappa=10$ fixed. The horizontal dotted lines represent $\Th=\kk/2\pi$, with $\kk$ of Eq.~\eqref{eq:temprel}, respectively for $|\alpha|=0,0.4,0.8$.}
\end{figure}

As can be seen in Fig.~\ref{fig:tom}, for all values of $|\alpha|<1$,
the temperature function $T_\om$ is almost constant in the low-frequency regime, until $\om$ approaches $\omm$,
where $\omm$ can be obtained from Eq.~\eqref{eq:ommax} through the replacement $D\mapsto D(1-\alpha)$.
Since the cutoff frequency $\omm$ is much larger than the various values of $\Teff$ (by at least a factor of 3), it is meaningful to say that five spectra in Fig.~\ref{fig:tom} are accurately Planckian.
In addition, in all cases, the effective temperature is well-approximated by the relativistic result $\Th=\kk/2\pi$.
However, in the limit $|\alpha|\to 1$, $\kk \to 0$, the deviations from Planckianity appear at lower and lower frequency. 
Then, for $\alpha>1$, the flow is everywhere subsonic, and there is no particle production because there are no negative norm modes with $\om > 0$~\cite{granada_proc}.
On the contrary, for $\alpha< - 1$, the flow is everywhere supersonic. A new critical frequency $\ommin<\omm$ appears, such that, for $0< \om<\ommin$, there now exists $4$ asymptotic in and out modes, two of them with positive norm and two with negative norm. Particle production will thus occur without sonic horizon, but the spectrum no longer diverges as $1/\om$ for $\om \to 0$~\cite{sf_thesis}.

Here, we consider only the case where the horizon is present ($|\alpha|<1$), such that the effective temperature $\Teff$ is well-defined for $\om\to0$. The steplike analysis can be generalized to this situation, yielding [see Eq.~\eqref{eq:Tstep}]
\begin{multline}
 \Tstep=\frac{3\sqrt{3}}{8}\,\omm(1+\alpha)\\ \times
\left[1-\left(q-\frac{1}{6}\right)(1-\alpha)D+O\left(D^2\right)\right],
\label{eq:T_step_as}
\end{multline}
where $\omm$ is again obtained from~Eq.~\eqref{eq:ommax} through $D\mapsto D(1-\alpha)$.
\begin{figure*}
 \includegraphics[width=0.85\columnwidth]{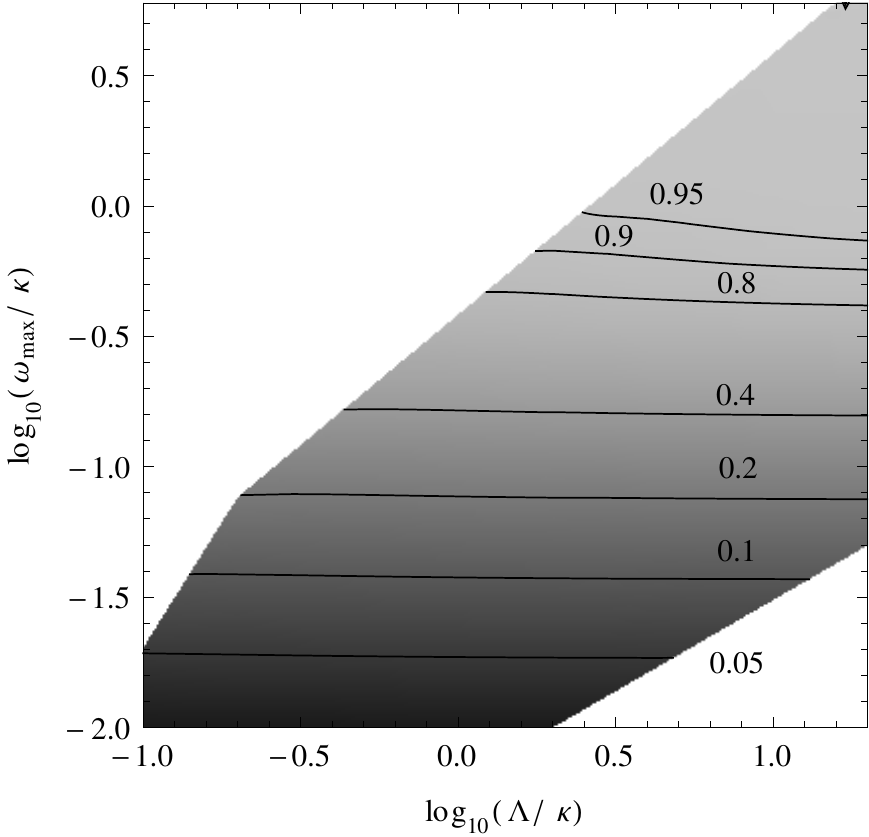}
 \hspace{2em}
 \includegraphics[width=0.85\columnwidth]{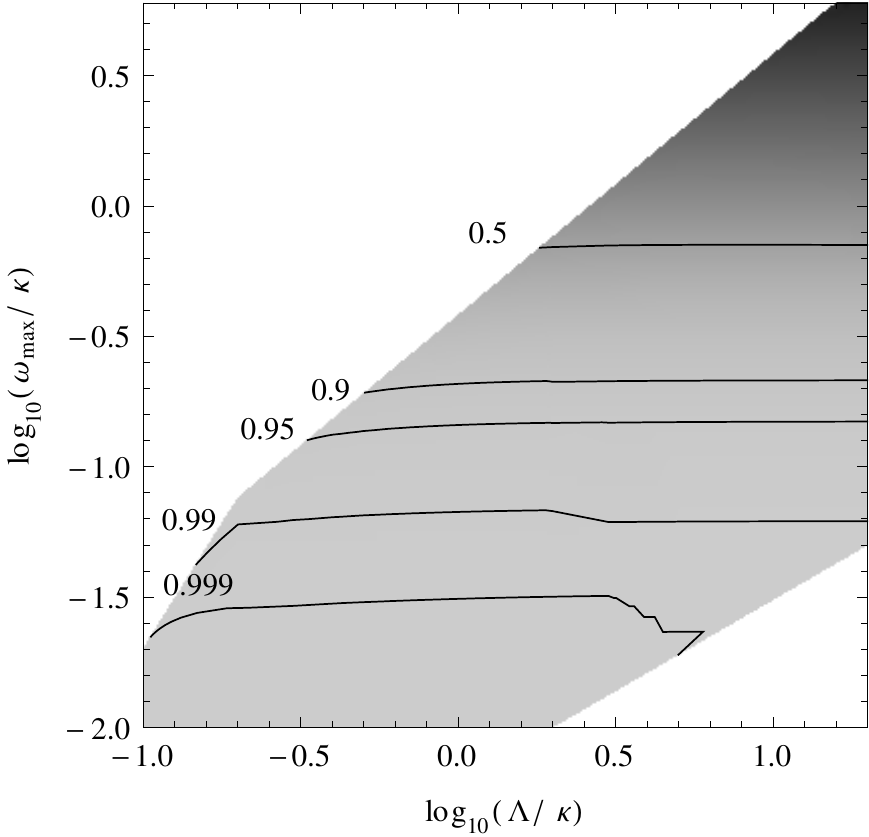}
 \caption{$\Teff/\Th$ (left panel) and $\Teff/\Tstep$ (right panel) as functions of $\Lambda/\ommax$ and $\ommax/\ka$ evaluated in the asymmetric flow profile of Eq.~\eqref{eq:velocity_as} for $\alpha =-0.5$.
The validity domains of both the Hawking and the steplike regime are not affected by the introduction of the parameter $\alpha$ which governs the asymmetry of the sub- and supersonic regions.}
 \label{fig:Tom2Dalpha}
\end{figure*}
Analogously, one can compute the effective temperature of left-going particles
\begin{multline}
 \Tvstep  =\frac{3\sqrt{3}}{2}\, \ommax(1+\alpha) \,
D^2  \left(q-\frac{1}{2}\right)^2\left[ 1 -\left(q-\frac{1}{6}\right)
\right.\\\left.\times
(1-\alpha)D
+ \alpha \,\frac{2q^2-2q+1}{2q-1}\,\frac{D}{2}
+O\left(D^2\right)\right].
\end{multline}
Since the first term in the expansions of $\Tvstep$ have the same dependence on $\alpha$ as the first term of $\Tstep$, their ratio
\begin{equation}
R_{\rm step}=4\left(q-\frac{1}{2}\right)^2 D^2\left[1+ \alpha \,\frac{2q^2-2q+1}{2q-1}\,\frac{D}{2}\right]+O\left(D^4\right)
\end{equation}
does not depend on $\alpha$ for small values of $D$.

In Fig.~\ref{fig:Tom2Dalpha}, $\Teff/\Th$ and $\Teff/\Tstep$ are plotted for a velocity profile with $\alpha=-0.5$.
Nothing qualitatively changes in these plots with respect to the simpler case with $\alpha=0$. 
Even if $\alpha$ is relevant in determining the temperature of the emitted spectrum  both in the Hawking and the steplike regimes, the transition between them is still  governed by $\ommax$ exactly as for $\alpha=0$.
Thus, it is again insensitive to the value of $\Lambda/\kappa$. We have extended our study for values of $\alpha$ closer than $-1$, and we observed no significant change.

\end{document}